\documentclass[final,3p,times,twocolumn]{elsarticle}

\usepackage{amssymb}
\usepackage{bm}
\usepackage{color} 
\usepackage{graphicx} %
\journal{Physica C}

\newcommand{\vecd}{\bm{d}}
\newcommand{\veck}{\bm{k}}
\newcommand{\vecs}{\bm{s}}
\newcommand{\imu}{{\rm i}}
\begin{document}

\begin{frontmatter}



\title{Theory of low-energy behaviors in topological $s$-wave pairing
 superconductors} 


\author{Y. Ota\corref{cor1}}

\author{Y. Nagai\corref{cor2}}
\author{M. Machida\corref{cor3}}

\address{CCSE, Japan Atomic Energy Agency, 178-4-4 Wakashiba, Kashiwa
 277-0871, Japan} 

\begin{abstract}
We construct a low-energy effective theory of topological $s$-wave
 pairing superconductors, focusing on the mean-field model of
superconductor $\mbox{Cu}_{x}\mbox{Bi}_{2}\mbox{Se}_{3}$. 
Our approach is second-order perturbation with respect to the inverse of the
 mass (i.e., large-mass expansion) in the Dirac-type electron dispersion
 from topological insulator $\mbox{Bi}_{2}\mbox{Se}_{3}$. 
Since the Dirac-type dispersion with a large mass describes
 non-relativistic electrons, the large-mass expansion corresponds to a
 low-energy theory with respect to the original setup. 
We show that the effective gap function has not only a $p$-wave-like 
component as the primary contribution, but also an $s$-wave-like one as
 higher-order corrections. 
The mixture of $p$- and $s$-wave explains the numerical
 results [Phys. Rev. B \textbf{89} (2014) 214506] of the non-magnetic
 impurity effects. 
\end{abstract} 


\end{frontmatter}


\section{Introduction}
Superconductor $\mbox{Cu}_{x}\mbox{Bi}_{2}\mbox{Se}_{3}$ and the related
compounds~\cite{Hor;Cava:2010,Wray;Hasan:2010,Shirasawa;Takahashi:2014,Sasaki;Ando:2014} attract a great deal of attention in condensed
matter physics since they are strong candidates for bulk topological
superconductors. 
To answer whether $\mbox{Cu}_{x}\mbox{Bi}_{2}\mbox{Se}_{3}$ is a genuine
topological superconductor, various experimental studies are performed,
including point-contact 
spectroscopy~\cite{Sasaki;Ando:2011,Kirzhner;Kanigel:2012},
magnetization curves~\cite{Das;Kadowaki:2011}, 
scanning tunneling spectroscopy~\cite{Levy;Stroscio:2013}, and
Knight-shift measurements~\cite{Hashimto;Tanaka:2012}. 

Impurity effects lead to a definite way to discriminating unconventional
features of superconductors~\cite{Balatsky;Zhu:2006}.  
The effects in $\mbox{Cu}_{x}\mbox{Bi}_{2}\mbox{Se}_{3}$ are of 
particular interest. 
Nagai \textit{et al.}~\cite{Nagai;Machida:2014} numerically studied
the non-magnetic impurity effects in the mean-field model of
$\mbox{Cu}_{x}\mbox{Bi}_{2}\mbox{Se}_{3}$~\cite{Sasaki;Ando:2011} (i.e.,
topological $s$-wave pairing superconductivity), by  
a self-consistent $T$-matrix approach. 
Since the model allows the presence of surface gapless
modes~\cite{Hao;Lee:2011}, one expects fragile
behaviors against non-magnetic impurities. 
In contrast to the intuitive assertion, the sensitivity is variable,
depending on the mass in the Dirac-type dispersion relation from
topological insulator $\mbox{Bi}_{2}\mbox{Se}_{3}$.  
When the mass is large, in-gap states in the density of states 
occurs; the superconducting state is not robust against non-magnetic
impurities. 
In contrast, when the mass is small, there is no in-gap state. 

In this paper, we study an effective theory to understand the
unconventional features of the mean-field model of
$\mbox{Cu}_{x}\mbox{Bi}_{2}\mbox{Se}_{3}$, motivated by the numerical
results by Nagai \textit{et al.}~\cite{Nagai;Machida:2014}. 
To answer how the robustness disappears depending on the mass
term, we derive a 
low-energy effective theory in a large-mass limit. 
Using the second-order perturbation with respect to the inverse of the
mass with a basis transformation to take higher-order corrections, we
show that the effective superconducting gap
function is described by a mixture of $p$- and $s$-wave-like components,
as seen in Eq.~(\ref{eq:eff_gap_full}). 
The latter is smaller than the former in the large-mass
limit. 
Therefore, we obtain the effective description of the
system, supporting the previous numerical calculations about the
non-magnetic impurity effects. 
The effective theory is useful for revealing the elementary
properties of the topological $s$-wave pairing superconductors, from a
gap-function-type point of view. 

\section{Model}
\label{sec:model}
The mean-field Hamiltonian is 
\(
\hat{H}_{\rm MF} 
=
(1/2) 
\sum_{\veck} 
\hat{\Psi}^{\dagger}_{\veck} 
\,
H(\veck)
\,
\hat{\Psi}_{\veck} 
\), 
with the Bogoliubov-de Gennes (BdG) Hamiltonian~\cite{Sasaki;Ando:2011}  
\begin{eqnarray}
&&
\hspace{-6.6mm}
H(\veck)
=
\frac{1 + \tau^{3}}{2} \otimes h_{0}(\veck)
+
\frac{1 - \tau^{3}}{2} \otimes [-h_{0}(-\veck)]^{\ast}
\nonumber 
\\
&&
\qquad
+
[
\tau^{+}
\otimes \Delta
+
({\rm h.c.})
] .
\label{eq:bdg_hamiltonian}
\end{eqnarray} 
The 8-component column vector $\hat{\Psi}_{\veck}$ has the electron
annihilation operators ($\hat{c}_{\veck,\sigma,s}$) in the upper 4-component 
block (particle space) and the electron creation ones 
($\hat{c}^{\dagger}_{-\veck,\sigma,s}$) in the lower 4-component block (hole
space), with momentum $\veck$, orbital $\sigma\,(=1,\,2)$, and spin
$s\,(=\uparrow,\,\downarrow)$.  
The $2\times 2$ Pauli matrices $\tau^{i}$ ($i=1,\,2,\,3$) represents the
Nambu space 
[$\tau^{+} = (\tau^{1} + \imu \tau^{2})/2$]. 
The superconducting pairing potential $\Delta$ has no
$\veck$-dependence. 
The model corresponds to a topological $s$-wave pairing
superconductor. 
Fu and Berg \cite{Fu;Berg:2010} proposed Eq.~(\ref{eq:bdg_hamiltonian}), 
based on short-range charge-density interaction . 

The normal-electron Hamiltonian in Eq.~(\ref{eq:bdg_hamiltonian}) is 
\begin{equation}
h_{0}(\veck) 
= 
-\mu 
+ 
M_{0} \gamma^{0} 
+
\sum_{i=1}^{3} d_{i}(\veck) \gamma^{0}\gamma^{i} ,
\label{eq:normal_hamiltonian}
\end{equation}
with chemical potential $\mu$ and spin-orbit couplings  
\(
(d_{1},\,d_{2},\,d_{3})
=
(
A k_{x},\, A k_{y},\, A^{\prime} k_{z}
)
\). 
The $4\times 4$ matrices $\gamma^{\mu}$ ($\mu=0,\,1,\,2,\,3$) satisfy the
Clifford algebra: 
\(
\gamma^{\mu}\gamma^{\nu} + \gamma^{\nu}\gamma^{\mu}
=
4 \eta^{\mu \nu}
\), 
with 
$\eta^{00}=1$, 
$\eta^{11}=-1$,
and
$\eta^{\mu \nu}=0$ for $\mu \neq \nu$.
They describe the orbit and the spin degrees of freedom in the system. 
Using the orbital $2\times 2$ Pauli matrices $\sigma^{i}$ and the spin
$2\times 2$ Pauli ones $s^{i}$, we have 
\(
\gamma^{0} = \sigma^{3} \otimes 1
\)
and 
\(
\gamma^{i} = \imu \sigma^{2} \otimes s^{i}
\). 
The role of $\gamma^{0}$ is of particular importance to the system
since it is related to spatial inversion; we find that 
\(
(\gamma^{0})^{\dagger} \, h_{0}(\veck) \, \gamma^{0} = h_{0}(-\veck)
\). 
Calculating the eigenvalues of $h_{0}$, we find that the normal
electrons have the massive Dirac-type dispersion relation, 
\(
\epsilon^{\pm}(\veck)
=
-\mu 
\pm 
[ M_{0}^{2} 
+ A^{2}(k_{x}^{2} + k_{y}^{2}) + A^{\prime\,2}k_{z}^{2}]^{1/2}
\). 
Without changing $\epsilon^{\pm}(\veck)$, the anisotropy along $z$-axis
in Eq.~(\ref{eq:normal_hamiltonian}) can be taken by a different
way~\cite{Fu;Berg:2010,Mizushima;Tanaka:2014}. 

We focus on the odd-parity fully-gapped superconducting order in this
paper. 
The pairing potential is 
\begin{equation}
\Delta
= 
\imu \Delta_{\rm odd} \imu \gamma^{0}\gamma^{2} 
=
\imu \Delta_{\rm odd} \sigma^{1} \otimes \imu s^{2} ,
\label{eq:pairing_potential}
\end{equation}
with a complex constant $\Delta_{\rm odd}$. 
Using the notations by Sasaki \textit{et al.}\cite{Sasaki;Ando:2011}, 
we find that 
\(
\Delta^{12}_{\uparrow \downarrow} 
= 
- \Delta^{12}_{\downarrow \uparrow}
=
\imu \Delta_{\rm odd}
\), 
\(
\Delta^{21}_{\uparrow \downarrow}
=
- \Delta^{21}_{\downarrow \uparrow}
=
\imu \Delta_{\rm odd}
\), 
and the other elements vanish. 
The transformation property of $\Delta$ with respect to the
spatial inversion is given by 
\(
(\gamma^{0})^{\dagger} \Delta (\gamma^{0})^{\ast}
=
-\Delta
\)
\cite{Sato:2010}. 
When the odd-parity gap appears, the BdG Hamiltonian does
not have a definite property under the spatial inversion. 
However, it has a symemtric character with respect to the combination of
the spatial inversion and rotation in the Nambu
space~\cite{Fu;Berg:2010,Sato:2010}; 
\begin{equation}
[( \tau^{3} \otimes 1 ) \Pi ]^{\dagger} 
H(\veck) 
[( \tau^{3} \otimes 1 ) \Pi ] = H(-\veck) ,
\label{eq:mod_tri_relation} 
\end{equation}
with
\(
\Pi
=
(1/2)(1 + \tau^{3}) \otimes \gamma^{0}
+ (1/2)(1 - \tau^{3}) \otimes (\gamma^{0})^{\ast}
\). 
This relation is essential for characterizing the model in terms of
invariants\,\cite{Fu;Berg:2010,Sato:2010}. 

A key point in our construction of a low-energy effective theory is to
focus on the mass term, $M_{0}\gamma^{0}$ in
Eq.~(\ref{eq:normal_hamiltonian})~\cite{Nagai;Machida:2014}.  
The use of a dimensionless quantity is convenient for our arguments, 
\begin{equation}
 \beta 
= 
\frac{A \bar{k}_{\rm F}}{|M_{0}|} 
=
\sqrt{
\frac{\mu^{2}}{M_{0}^{2}}
-1
},
\end{equation}
with the Fermi momentum  
\(
\bar{k}_{\rm F} = (1/A) (\mu^{2} - M_{0}^{2})^{1/2}
\). 
When $\beta$ is small, the system is in a large-mass (non-relativistic)
region. 
The Fermi surafce is determiend by a larger eigenvalue of
Eq.~(\ref{eq:normal_hamiltonian}), i.e., 
\(
0 = \epsilon^{+} (\bar{k}_{\rm F})
\), with $\mu > |M_{0}|$. 
We obtain $\bar{k}_{\rm F}$ for the spherical Fermi
surface, redefining the momentum; 
$(A^{\prime} /A) k_{z} \to k_{z}$. 

Focuing on $\beta$ is effective for understanding the system. 
The stable superconducting order is ruled by the value of 
$(1 + \beta^{2})^{-1/2}\,(=|M_{0}|/\mu)$, within the linearized gap
equation~\cite{Fu;Berg:2010}. 
The value of $\beta$ is also relevant to the response properties to
impurities of an odd-parity superconducting gap with a point
node~\cite{Nagai;1410.4646}. 

Before closing this section, we sumamrize the properties of
Eq.~(\ref{eq:bdg_hamiltonian}) when $M_{0}=0$ (i.e., 
$\beta \to \infty$). 
Using the chiral symmetry~\cite{Fu;Berg:2010}, we find that 
\(
\gamma^{5} h_{0} - h_{0} \gamma^{5} =0
\)
and 
\(
\gamma^{5} \Delta - \Delta \gamma^{5} = 0
\), 
with 
\(
\gamma^{5} 
=
\imu \gamma^{0}\gamma^{1}\gamma^{2}\gamma^{3}
=
\sigma^{1} \otimes 1
\). 
Hence, exchanging the order between $\tau^{i}$ (Nambu) and
$\sigma^{i}$ (orbital), we have 
\(
H(\veck)|_{M_{0}=0}
=
[(1 + \sigma^{1})/2]\otimes H^{\rm R}(\veck)
+
[(1 - \sigma^{1})/2]\otimes H^{\rm L}(\veck)
\), 
with 
\begin{eqnarray}
&&
\hspace{-6.6mm}
H^{({\rm R,L})}(\veck)
=
\frac{1 + \tau^{3}}{2} 
\otimes 
[
- \mu + \kappa_{({\rm R,L})} A (\bar{\veck} \cdot \vecs)
]
\nonumber \\
&&
\qquad\quad\,
+
\frac{1 - \tau^{3}}{2} 
\otimes 
[
\mu + \kappa_{({\rm R,L})} A(\bar{\veck} \cdot \vecs)^{\ast}
]
\nonumber \\
&&
\qquad\quad\,
+
[\tau^{+} \otimes 
\kappa_{({\rm R,L})} (\imu \Delta_{\rm odd}) 
\imu s^{2} + ({\rm h.c.})] ,
\label{eq:massless_h}
\end{eqnarray}
where
$\bar{\veck} = [k_{x},\,k_{y},\,(A^{\prime}/A)k_{z}]$
and 
$(\kappa_{\rm R},\,\kappa_{\rm L})=(1,\,-1)$. 
In each block, the normal part is transformed into a diagonal
form by 
\(
[(1 + \tau^{3})/2] \otimes U_{\veck}
+
[(1 - \tau^{3})/2] \otimes (-\imu s^{2} U_{\veck})
\), 
with 
\(
U^{\dagger}_{\veck} (\bar{\veck}\cdot \vecs) U_{\veck} 
=
|\bar{\veck}| \, {\rm diag}(1,\,-1) 
\). 
To show this statement, a relation of the Pauli matrices is
employed; 
\(
s^{2} \vecs s^{2} = -\vecs^{\ast} 
\). 
After the basis transformation, 
the superconducting part is independent of $\veck$ 
since  
\(
U^{\dagger}_{\veck}\, \imu s^{2}\, (-\imu s^{2}U_{\veck}) = 1
\). 
We mention that the system with $M_{0}=0$ is different from the
conventional $s$-wave superconductors; 
the odd-parity property of the pairing potential appears as a sign
change between the right- and left-handed
blocks~\cite{Nagai;Machida:2014}. 
However, the gap function is considered to be that of an $s$-wave state
as long as the BdG Hamiltonian is decoupled into the blocks with
different kinds of chirality. 
Therefore, we may say that the system with $M_{0}=0$ reduces to an
$s$-wave superconducting model~\cite{Nagai;Machida:2014,Fu;Berg:2010}. 

The character of this $s$-wave-like gap function implies that the
odd-parity state with a large $\beta$ is robust against non-magnetic
impurities. 
Nagai \textit{et al.}~\cite{Nagai;Machida:2014} numerically
found no occurrence of in-gap states under
non-magnetic impurities when $\beta > 1$. 
Michaeli and Fu~\cite{Michaeli;Fu:2012} proposed the protection of the
odd-parity superconducting state against non-magnetic impurities in
terms of spin-orbit locking effects. 
The effects are predominant when $\beta$ (i.e., $A$) is large; in a
relativistic region this mechanism is reasonable. 

\section{Results: Derivation of low-energy effective theory}
\label{sec:results}
Now, we construct a low-energy effective theory when $\beta \sim 0$ 
(i.e., large-mass expansion).  
Our approach is similar to the arguments in 
semiconductor-superconductor junction
systems~\cite{Alicea;Fisher:2011}, but we take higher-order
corrections. 
The corrections would be primary, taking a large $\beta$. 
The arguments at the end of Section \ref{sec:model} show that an 
$s$-wave character is manifest in the system when $\beta \to \infty$. 
Thus, our approach is useful for understanding how this $s$-wave
character disappears when $\beta \to 0$. 
Throughout this section, we exchange the order between the Nambu space
and the orbital degrees of freedom.

\subsection{Effective theory without higher-order corrections}
\label{subsec:effective_wo}
We show the low-energy effective theory without higher-order
corrections~\cite{Nagai;Machida:2014}.  
The arguments can straightforwardly extend to the case with
corrections. 
 
When $\beta \simeq 0$, the mass term in Eq.~(\ref{eq:normal_hamiltonian}) is
predominant. 
Let us seek the perturbation terms in Eq.~(\ref{eq:bdg_hamiltonian}). 
In the normal part, the spin-orbit couplings are regarded as
the perturbation since the magnitude of these terms is characterized by 
$A \bar{k}_{\rm F}$. 
Moreover, in a weak-coupling superconductor, the contributions from
the pairing potential are smaller than the ones from the normal
electrons. 
Thus, the paring potential in the mean-field Hamiltonian should be the
perturbation. 
Therefore, we find that $H(\veck) = H_{0}(\veck) + V(\veck)$, with the
free part 
\(
H_{0} 
= 
[(1+\sigma^{3})/2] \otimes H_{0}^{+} 
+
[(1-\sigma^{3})/2] \otimes H_{0}^{-}  
\)
and the perturbation 
\(
V 
=
\sigma^{1} \otimes V^{+-}
\), 
where
\begin{eqnarray}
&&
\hspace{-6.6mm}
H_{0}^{\pm}(\veck)
=
(-\mu \pm M_{0} ) \tau^{3} \otimes 1, 
\label{eq:bare_free}
\\
&&
\hspace{-6.6mm}
V^{+-}(\veck)
=
\frac{1+\tau^{3}}{2} \otimes A (\bar{\veck}\cdot \vecs)
+
\frac{1-\tau^{3}}{2} \otimes A (\bar{\veck}\cdot \vecs)^{\ast}
\nonumber 
\\
&&
\qquad\quad\,
+
[\tau^{+} \otimes (\imu \Delta_{\rm odd}) \imu s^{2} 
+ ({\rm h.c.})]. 
\label{eq:bare_perturbation}
\end{eqnarray}
Since the perturbation contains the spin-orbit couplings and the pairing
potential, the perturbation expansion is valid within 
$|M_{0}| \gg A \bar{k}_{\rm F},\, |\Delta_{\rm odd}|$. 

To perform the perturbation systematically, we use the orthogonal
projectors $\mathcal{P}$ and $\mathcal{Q}$, defined as 
\begin{equation}
\mathcal{P} = \frac{1 + \sigma^{3}}{2} \otimes 1 \otimes 1,
\quad
\mathcal{Q} = \frac{1 - \sigma^{3}}{2} \otimes 1 \otimes 1. 
\label{eq:projector}
\end{equation}
Throughout this section, we focus on a postive $M_{0}$.  
In this case, the subspace given by $\mathcal{P}$ is our target space. 
When $M_{0}$ is negative, we can dervie the effective theory exchanging
the role between $\mathcal{P}$ and $\mathcal{Q}$. 
Using the second-order Brillouin-Wigner perturbation
approach~\cite{Hubac;Wilson:2010}, we obtain the effective Hamiltonian, 
\begin{eqnarray}
&&
\hspace{-6.6mm}
H_{\rm eff}(\veck)
=
\mathcal{P} H_{0}(\veck) \mathcal{P}
\nonumber \\
&&
\qquad\quad\,
+
\sum_{m=\pm}
[
\mathcal{P} V(\veck)
\mathcal{Q}
] 
R^{m}(\veck)
[
\mathcal{Q} V(\veck) \mathcal{P}
],
\label{eq:BW_expansion}
\end{eqnarray} 
with 
\(
( E^{m}_{0} - \mathcal{Q}H_{0}\mathcal{Q} ) R^{m} = \mathcal{Q}
\)
and 
\(
E^{\pm}_{0} = \pm (-\mu + M_{0})
\). 
A similar technique is applied to deriving a low-energy effective theory
in a different model of superconducting topological
insulator~\cite{Hao;Lee:2015}. 
After straightforward calculations, we find that 
\(
H_{\rm eff}(\veck)
=
[(1+\sigma^{3})/2] \otimes H_{\rm eff}^{+}(\veck)
\), 
where
\begin{eqnarray}
&&
\hspace{-6.6mm}
H_{\rm eff}^{+}(\veck)
=
\frac{1 + \tau^{3}}{2} \otimes h_{{\rm eff},0}(\veck)
\nonumber \\
&&
\qquad\quad \,
+
\frac{1 - \tau^{3}}{2} \otimes [-h_{{\rm eff},0}(-\veck)]^{\ast} 
\nonumber \\
&&
\qquad\quad \,
+
[
\tau^{+} \otimes \Delta_{{\rm eff},0}(\veck) + ({\rm h.c.})
] .
\label{eq:eff_h_wo}
\end{eqnarray} 
In the vicinity of the Fermi surface (i.e., $\mu \approx M_{0}$), we
obtain 
\(
h_{{\rm eff},0}(\veck)
\approx 
(1/2M_{0})\, 2 [
(A \bar{\veck})^{2}
-
|\Delta_{\rm odd}|^{2} ]
\) 
and 
\begin{equation}
\Delta_{{\rm eff},0}(\veck)
\approx 
2 \beta (\imu \Delta_{\rm odd}) \,
[\bar{\vecd}(\veck) \cdot \vecs]\, \imu s^{2},
\label{eq:eff_gap_wo}
\end{equation}
with 
\(
\bar{\vecd} = \bar{\veck} / \bar{k}_{\rm F}
\). 
Thus, we find that the effective gap function corresponds to a
$p$-wave-like state~\cite{Nagai;Machida:2014}. 

\subsection{Mixture of $p$- and $s$-wave components}
\label{subsec:mixture}
Let us show the low-energy effective theory with higher-order
corrections. 
We take a positive $M_{0}$ again; the target subspace is specified by
$\mathcal{P}$ in Eq.~(\ref{eq:projector}). 
A straightforward way to taking the corrections is to add higher-order
expansion terms to Eq.~(\ref{eq:BW_expansion}), but this approach would be
messy. 
We use an alternative way, i.e., a second-order perturbation approach
with a basis transformation.
Let us concisely summarize our approach. 
A unitary transformation is first applied to
Eq.~(\ref{eq:bdg_hamiltonian}), to obtain a better basis in perturbation. 
Then, in this transformed basis the second-order perturbation equivalent
to that in Section ~\ref{subsec:effective_wo} is performed. 
Thus, the corrections from the basis transformation would appear in
Eq.~(\ref{eq:eff_h_wo}).  

Using a basis transformation, the BdG Hamiltonian is partially
diagonalized. 
The perturbation in the transformed basis would lead to better
expansion. 
To focus on the pairing potential leads to a clue of finding a proper
basis.  
We find that this term commutes with the mass term in
Eq.~(\ref{eq:bdg_hamiltonian}). 
The proof is done by the same technique of showing relation
(\ref{eq:mod_tri_relation}). 
Thus, the basis transformation should include rotation in the Nambu
space. 
We propose the basis transformation given by a unitary matrix with a small
angle $\beta$, 
\begin{equation}
\mathcal{S}_{\beta} 
=
\exp \left\{
\frac{\imu \beta}{2}
\sigma^{3}
\otimes
[
\chi_{\rm odd}
\tau^{+}
+
({\rm h.c.})
]
\otimes
s^{2}
\right\} ,
\label{eq:tansformation}
\end{equation}  
with 
\(
\chi_{\rm odd}
=
-\Delta_{\rm odd}/|\Delta_{\rm odd}|
\). 
We denote the unitary-transformed BdG Hamiltonian as 
\(
H_{\beta}(\veck)
\). 
The free part and the perturbation are, respectively, expressed by
\(
H_{0,\, \beta}(\veck)
=
\mathcal{S}_{\beta} 
H_{0}(\veck) 
\mathcal{S}^{\dagger}_{\beta}
\)
and
\(
V_{\beta} (\veck)
=
\mathcal{S}_{\beta} 
V(\veck) 
\mathcal{S}^{\dagger}_{\beta}
\). 
To perform small-$\beta$ expansion systematically in the
unitary-transformed formulae, we focus on an algebraic relation in
Eq.~(\ref{eq:tansformation}). 
Let us rewrite 
\(
\mathcal{S}_{\beta} = \exp [(-\imu \beta /2) W]
\), 
with 
\(
W = -\sigma^{3}
\otimes
[
\chi_{\rm odd}
\tau^{+}
+
({\rm h.c.})
]
\otimes
s^{2}
\). 
We find that $W$ is a Hermite matrix and does not contain any
small parameters (i.e., 
$\beta$ and $|\Delta_{\rm odd}|/|M_{0}|$). 
Since $\chi_{\rm odd} = - \Delta_{\rm odd}/|\Delta_{\rm odd}|$, we have 
\(
W^{2} = 1
\). 
Therefore, we find that
\(
\mathcal{S}_{\beta} 
= 
\cos[(\beta /2)]  - \imu W \sin[(\beta /2)]
=
1 - ( \imu \beta /2 ) W + \mathcal{O}(\beta^{2}) 
\). 
It indicates that for a matrix $M$ the unitary-transformed one is 
\(
M_{\beta} = M + (\imu \beta /2)[M,\,W] + \mathcal{O}(\beta^{2})
\). 

We explicitly write down the formulae of 
\(
H_{0,\, \beta}(\veck)
\)
and 
\(
V_{\beta} (\veck)
\). 
First, we focus on the free part. 
Using a similar manner to Eq.~(\ref{eq:bare_free}), we obtain
\(
H_{0,\,\beta}
=
[(1+\sigma^{3})/2] \otimes H^{+}_{0,\,\beta}
+
[(1-\sigma^{3})/2] \otimes H^{-}_{0,\,\beta}
\). 
We find that 
\begin{eqnarray}
&&
\hspace{-6.6mm}
H^{\pm}_{0,\,\beta}(\veck)
=
(-\mu \pm M_{0})\tau^{3} \otimes 1
\nonumber 
\\
&&
\quad \,\,\,
+
\beta
[
\Delta^{(\pm)}(\veck) \tau^{+} \otimes \imu s^{2}
+
({\rm h.c.})
] 
+ \mathcal{O}(\beta^{2}),
\label{eq:transformed_normal}
\end{eqnarray}
with 
\(
\Delta^{(\pm)}(\veck)
=
-\imu (-\mu \pm M_{0}) \chi_{\rm odd}
\). 
We find that $\Delta^{(+)}(\veck)$ vanishes in the vicinity
of the Fermi surface since $\mu \approx M_{0}$. 
As a result, we consider
that 
\(
H^{+}_{0,\,\beta} \approx H^{+}_{0}
\) in the perturbation expansion. 
From this point of view, the present basis transformation 
\textit{minimally} deforms the original issue. 
Next, we examine the transformed perturbation. 
We find that 
\begin{eqnarray}
&&
\hspace{-6.6mm}
V_{\beta}(\veck)
=
\sigma^{1} \otimes V^{+-}(\veck) 
\nonumber \\
&&
\quad \,\,\,
+
(-\beta)|\Delta_{\rm odd}|
\sigma^{2} \otimes 1 \otimes 1 
+ \mathcal{O}(\beta^{2}).
\end{eqnarray}
The second term is an important effect caused by the basis
 transformation. 
This term is relevant to an $s$-wave behavior in the
large-mass limit.  

Now, we show the low-energy effective Hamiltonian. 
We repeat the same discussion as in Section \ref{subsec:effective_wo},
with the free part $H_{0,\,\beta}$ and the perturbation $V_{\beta}$. 
The calculation details are shown in \ref{append:details}. 
Performing the second-order perturbation expansion, we obtain the
effective Hamiltonian in the subspace specified by $\mathcal{P}$. 
The Hamitlnian $H^{+}_{\rm eff}(\veck)$ in this subspace is attainable,
replacing $h_{{\rm eff},\,0}$ and $\Delta_{{\rm eff},\,0}$ in
Eq.~(\ref{eq:eff_h_wo}), respectively, with 
\(
h_{{\rm eff},\,\beta} 
\)
and
\(
\Delta_{{\rm eff},\,\beta}
\). 
In the vinicity of the Fermi surface, we find that 
\(
h_{{\rm eff},\,\beta}(\veck)
\approx 
h_{{\rm eff},\,0}(\veck)
\). 
The effective gap is
\begin{equation}
\Delta_{{\rm eff},\,\beta}(\veck)
\approx
2\beta (\imu \Delta_{\rm odd})
\left(
\bar{\vecd}(\veck) \cdot \vecs
+
\imu
\frac{|\Delta_{\rm odd}|}{M_{0}} 
\right)\, \imu s^{2} .
\label{eq:eff_gap_full}
\end{equation}
The first term is equal to Eq.~(\ref{eq:eff_gap_wo}) and describes 
a $p$-wave-like state, whereas the second term is an $s$-wave-like
state. 
The physical origin of the $p$-wave component is a cross term between
the spin-orbit couplings and the superconducting gap. 
The second term comes from a higher-order correction in the perturbation
term derived by the basis transformation. 
Thus, in a large-mass limit, the system is effectively expressed by a 
mixture of the $p$-wave and the $s$-wave components. 
The primary contribution is the $p$-wave one in this limit. 
However, when the second term is not negligible, the system can behaves as
an $s$-wave superconducting state. 

\subsection{Link of effective theory with odd-parity pairing potential}
We argue a link of the effective gap with the
pairing potential given by Eq.~(\ref{eq:pairing_potential}), from a
parity point of view. 
One approach is to make a transformation of spatial inversion in the
projected subspace appeared in the perturbation analysis. 
The resultant formula might lead to a transformation property of
$\Delta_{{\rm eff},\,\beta}$, similar to that of $\Delta$ in Section
\ref{sec:model}. 
For this purpose, we implement the projector $\mathcal{P}$ on
the spatial-inversion transformation in the full BdG formulation. 
According to the calculations in \ref{append:spatial_inversion}, we 
find that the projected spatial inversion is represented by the
identity matrix.  
Thus, this transformation leads to no information on the
transformation property of the effective gap. 

We take an alternative approach of finding a connection of the effective
theory with the full BdG formulation. 
Let us compare the effective theory with that in the other projected
subspace. 
We change a kind of the projectors: 
\(
\mathcal{P} = [(1 - \sigma^{3})/2] \otimes 1 \otimes 1
\)
and
\(
\mathcal{Q} = [(1 + \sigma^{3})/2] \otimes 1 \otimes 1
\). 
We have two setups relevant to this choice. 
One is the case of $\mu < 0$ with $M_{0}> 0$, while the other is that of
$\mu > 0$ with $M_{0} < 0$. 
We take the former in this paper. 
This setup means that the chemical potential intersets
with a lower energy band of the normal-electron Dirac dispersions; the
Fermi surface is defined by 
\(
0 = \epsilon^{-}(\veck)
\), 
leading to $\mu = - M_{0} + \mathcal{O}(\beta^{2})$. 
The derivation of the corresponding effective theory is parallel to the
previous case. 
The calculations are shown in \ref{append:details}. 
We obtain the effective gap 
\begin{equation}
\Delta_{{\rm eff},\,\beta}(\veck) 
\approx
- 2\beta (\imu \Delta_{\rm odd})
\left(
\bar{\vecd}(\veck) \cdot \vecs
-
\imu
\frac{|\Delta_{\rm odd}|}{M_{0}} 
\right)\, \imu s^{2} .
\label{eq:eff_gap_full_another}
\end{equation}
We note that the other setup (i.e., $\mu > 0$ and $M_{0}<0$) leads to
Eq.~(\ref{eq:eff_gap_full_another}), up to an overall phase. 
We find again that the first term has momentum dependence (i.e.,
$p$-wave-like component), while the second term is independent of
momentum (i.e., $s$-wave-like component). 
The difference from Eq.~(\ref{eq:eff_gap_full}) is the relative sign
between the two components, except for a global phase. 
The fact that the spatial inversion in the full BdG formalism is ruled
by the orbital-space Pauli matrix $\sigma^{3}$ causes the discrepancy of
the effective gaps. 

The above arguments are contrast to an effective theory with an
even-parity (i.e, conventional $s$-wave) pairing potential. 
An even-parity gap can be written
by~\cite{Nagai;Machida:2014} 
\begin{equation}
\Delta 
= 
\Delta_{\rm even} \imu \gamma^{0}\gamma^{5} \imu \gamma^{2}
=
-\imu \Delta_{\rm even} 1 \otimes \imu s^{2} ,
\end{equation}
with a complex constant $\Delta_{\rm even}$. 
We find that in the orbital space a trivial (i.e., identity) matrix 
appears. 
The low-energy effective theory with $\beta \to 0$ is attainbale in
a similar way to that of the odd-parity state~\cite{Nagai;Machida:2014}. 
One can find that the effective gap contains only a $\veck$-independent
(i.e., s-wave-like) component.
Moreover, the effective gap with
$\mathcal{P}=[(1+\sigma^{3})/2]\otimes 1 \otimes 1$ is equal to that
with $\mathcal{P}=[(1-\sigma^{3})/2]\otimes 1 \otimes 1$ since the
even-parity pairing potential is invariant under any
transformation in the orbital space. 
To sum up, the odd-parity property of the pairing potential in the
full BdG Hamiltonian appears as an inter-component sign difference in the
effetive gap depending on a kind of the projectors into a low-energy
space. 

\section{Conclusion}
We built up a low-energy effective theory, focusing on a model of
superconductor $\mbox{Cu}_{x}\mbox{Bi}_{2}\mbox{Se}_{3}$, motivated by
the numerical results of the non-magnetic impurity effects. 
Using the second-order perturbation with respect to the mass in the
Dirac-type electron dispersion and performing a basis transformation to
take higher-order corrections, we showed that in the low-energy
effective theory the effective superconducting gap is described by a 
mixture of a $p$-wave component and an $s$-wave component. 
We stress that the latter is smaller than the former in the large-mass
limit. 
Thus, we obtained the effective description of the system, supporting
the previous numerical calculations about the non-magnetic impurity
effects. 
An interesting future work is to clarigy how the $p$- and $s$-wave
components in the effective gap contribute to the topological invariant
in this model. 

\section*{Acknowledgements}
We would like to thank H. Nakamura and K. Kobayashi for their helpful
discussion. 
This study is partially supported by Grant-in-Aid for Scientific
Research (S) Grant Number 23226019, and JSPS KAKENHI Grant Numbers
24340079 and 26800197. 

\appendix
\section{Details of perturbation calculations}
\label{append:details}
We show the perturbation calculations to derive the low-energy effective
theory, within $M_{0} > 0$. 
In the Appendix we use the same arrangement of matrices as
that in Section \ref{sec:results}; the orbital space is first written, 
next the Nambu space is put, and finally the spin space is placed. 
We focus on the case of performing basis transformation
(\ref{eq:tansformation}). 
We are going to calculate 
\begin{equation}
H_{1,\,\beta} (\veck)
=
\sum_{m=\pm}
[
\mathcal{P} V_{\beta} (\veck)
\mathcal{Q}
] 
R^{m}_{\beta}(\veck)
[
\mathcal{Q} V_{\beta}(\veck) \mathcal{P}
].
\label{eq:corrections_transformed}
\end{equation}
The subscript $\beta$ indicates that our calculations are performed
after the basis transformation given by Eq.~(\ref{eq:tansformation}). 
When $\beta=0$, this expression is equal to the second term in
Eq.~(\ref{eq:BW_expansion}). 
The intermediate matrix $R^{m}_{\beta}$ represents the propergator
via virtual states in the perturbation expansion, given by 
\(
(E_{0,\,\beta}^{m} - \mathcal{Q} H_{0,\,\beta} \mathcal{Q})
R^{m}_{\beta} = \mathcal{Q}
\). 
Two distinct eigenvalues of $\mathcal{P} H_{0,\,\beta} \mathcal{P}$ are
denoted by $E_{0,\,\beta}^{m}$. 
Each of them is four-fold degenerate. 
Each matrix element of 
\(
\mathcal{P} V_{\beta} (\veck) \mathcal{Q} 
\) 
is regarded as transition amplitude from a subspace defined by
$\mathcal{Q}$ to that by $\mathcal{P}$. 

We argue a way of constructing $R_{\beta}^{m}$. 
First, we study the case when $\mu$ is positive. 
The Fermi surface is defined by $0=\epsilon^{+}(\veck)$. 
This is the same setting as that in the main text. 
The projectors $\mathcal{P}$ and $\mathcal{Q}$ are given by
Eq.~(\ref{eq:projector}).  
According to the arguments on Eq.~(\ref{eq:transformed_normal}) in
Section \ref{subsec:mixture}, we find that  
\(
\mathcal{P} H_{0,\,\beta} \mathcal{P}
=
[(1+\sigma^{3})/2] \otimes H_{0,\,\beta}^{+}
=
\mathcal{P} H_{0,\,\beta=0} \mathcal{P}
+
\mathcal{O}(\beta^{2})
\) 
since 
\(
\Delta^{+}(\veck) 
= \mathcal{O}(\beta^{2})
\). 
Thus, we obtain 
\(
E_{0,\,\beta}^{m} 
= E_{0,\,\beta=0}^{m} + \mathcal{O}(\beta^{2})
\). 
Moreover, we find in Eq.~(\ref{eq:transformed_normal}) that 
\(
\Delta^{-} (\veck) = \imu 2M_{0} \chi_{\rm odd} +
\mathcal{O}(\beta^{2})
\). 
Fully diagonalizing $\mathcal{Q} H_{0,\,\beta} \mathcal{Q}$, we can find
that the changes of the spectrum (and the eigenvectors) from the formula
with $\beta=0$ are in order of $\beta^{2}$. 
Integrating these arguments, we show that 
\(
R_{\beta}^{m} = R_{\beta=0}^{m} + \mathcal{O}(\beta^{2})
\). 
Then, we have
\begin{equation}
R_{\beta}^{m}\, |_{\mu > 0}
=
\frac{1}{2M_{0}} 
\frac{1-\sigma^{3}}{2} 
\otimes \tau^{3}
\otimes 1 
+
\mathcal{O}(\beta^{2}) .
\end{equation}
Second, we examine $R_{\beta}^{m}$ when $\mu$ is negative. 
This setting corresponds to the case when the Fermi surface intersets
with a lower energy of the normal Hamiltonian; 
\(
0 = \epsilon^{-}(\veck)
\), 
indicating that 
\(
\mu = - M_{0} + \mathcal{O}(\beta^{2})
\). 
Thus, the predominant part of the normal Hamiltonian comes from the
subspace characterized by the negative eigenvalue of $\sigma^{3}$. 
Therefore, the projectors in Eq.~(\ref{eq:corrections_transformed}) need
to be set by
\(
\mathcal{P} = [(1-\sigma^{3})/2] \otimes 1 \otimes 1
\)
and
\(
\mathcal{Q} = [(1+\sigma^{3})/2] \otimes 1 \otimes 1
\). 
The arguments in $\mu > 0$ with swapping from the role of $\mathcal{P}$
to that of $\mathcal{Q}$ lead to 
\begin{equation}
R_{\beta}^{m}\, |_{\mu < 0}
=
\frac{-1}{2M_{0}} 
\frac{1+\sigma^{3}}{2} 
\otimes \tau^{3}
\otimes 1 
+
\mathcal{O}(\beta^{2}) .
\end{equation}

The calculations of $\mathcal{P} V_{\beta} (\veck) \mathcal{Q}$ 
are straightforwardly performed, using the algebraic properties of the
orbital-space Puali matrices. 
When $\mu > 0$, we have
\begin{equation}
\mathcal{P} V_{\beta} \mathcal{Q} \, |_{\mu > 0}
=
\sigma^{+} \otimes 
(
V^{+-} -\imu \phi 1 \otimes 1
)
+
\mathcal{O}(\beta^{2})  ,
\label{eq:trans_amp_pos_mu}
\end{equation} 
with 
\(
\sigma^{+} = (\sigma^{1} + \imu \sigma^{2})/2 
\) 
and 
\(
\phi = (-\beta) |\Delta_{\rm odd}|
\). 
In contrast, when $\mu < 0$, the formula is 
\begin{equation}
\mathcal{P} V_{\beta} \mathcal{Q} \,|_{\mu < 0}
=
\sigma^{-} \otimes 
(
V^{+-} + \imu \phi 1 \otimes 1
)
+
\mathcal{O}(\beta^{2}) ,
\label{eq:trans_amp_neg_mu}
\end{equation} 
with $\sigma^{-} = (\sigma^{+})^{\dagger}$. 
It is worth for pointing out the relative sign between the first and
second terms in part of the Nambu-spin space (i.e., inside the
parentheses). 
The difference of the signs between Eqs.~(\ref{eq:trans_amp_pos_mu}) and
(\ref{eq:trans_amp_neg_mu}) comes from that of the projectors. 
Thus, a character in the orbital space alters the transition amplitude
in the perturbation expansion. 

Now, we show the explicit formulae of $H_{1,\,\beta}$, using the
expressions of $R_{\beta}^{m}$ and $\mathcal{P}V_{\beta}\mathcal{Q}$. 
When $\mu$ is positive, we find that
\begin{eqnarray}
&&
\hspace{-6.6mm}
H_{1,\,\beta}\,|_{\mu > 0}
=
\frac{1 + \sigma^{3}}{2}
\otimes 
\frac{1}{M_{0}} 
\big[
V^{+-} (\tau^{3}\otimes 1 ) V^{+-}
\nonumber \\
&&
\qquad
-\imu \phi [\tau^{3} \otimes 1,\, V^{+-}]
\big] + \mathcal{O}(\beta^{2}) . 
\end{eqnarray} 
In contrast, when $\mu$ is negative, we have
\begin{eqnarray}
&&
\hspace{-6.6mm}
H_{1,\,\beta}\, |_{\mu < 0}
=
\frac{1 - \sigma^{3}}{2}
\otimes 
\frac{-1}{M_{0}} 
\big[
V^{+-} (\tau^{3}\otimes 1 ) V^{+-}
\nonumber \\
&&
\qquad
+ \imu \phi [\tau^{3} \otimes 1,\, V^{+-}]
\big] + \mathcal{O}(\beta^{2}) . 
\end{eqnarray} 
In part of the Nambu-spin space, the first term, 
\(
V^{+-} (\tau^{3}\otimes 1 ) V^{+-}
\) 
contains an effective gap with momentum dependence, whereas the second
term, 
\(
\mp \imu \phi  [\tau^{3} \otimes 1,\, V^{+-}]
\)
leads to an effective gap without momentum dependence. 

\section{Spatial inversion in a projected space}
\label{append:spatial_inversion}
We show an approach of formulating spatial inversion in a projected space. 
We take the projected space specified by 
$[( 1 + \sigma^{3} )/2] \otimes 1 \otimes 1$. 
In the Appendix we use the same arrangement of matrices as
that in Section \ref{sec:results}. 

The transformation matrix of spatial inversion in the Nambu space is
expressed by 
\(
\Pi = \sigma^{3} \otimes 1 \otimes 1
\) 
[See the text below Eq.~(\ref{eq:mod_tri_relation})]. 
We find that the spatial inversion leads to 
\(
\Pi^{\dagger} O \Pi 
\) for a matrix $O$ in the Nambu space, such as the BdG Hamiltonian. 
We first perform the basis transformation given by
Eq.~(\ref{eq:tansformation}) on $\Pi$: 
\(
\Pi_{\beta} = \mathcal{S}_{\beta} \Pi \mathcal{S}_{\beta}^{\dagger}
\). 
We write 
\(
\mathcal{S}_{\beta} = \exp [ (-\imu \beta /2) W]
\), similar to the arguments on Eq.~(\ref{eq:tansformation}) in Section
\ref{subsec:mixture}.  
Since $[\Pi,\, W] = 0$, this transformation has no effect on
$\Pi$; $\Pi_{\beta} = \Pi$. 
Next, we expand $\Pi$ in terms of the projectors given by
Eq.~(\ref{eq:projector}), using the resolution of unity, 
\(
\mathcal{P} + \mathcal{Q} = 1 \otimes 1 \otimes 1
\). 
Since $[\Pi,\, \mathcal{P}] = [\Pi,\, \mathcal{Q}] = 0$ and
$\mathcal{P}\mathcal{Q}=\mathcal{Q}\mathcal{P}=0$, we obtain 
\(
\Pi_{\beta} 
= 
\mathcal{P} \Pi_{\beta} \mathcal{P}
+
\mathcal{Q} \Pi_{\beta} \mathcal{Q}
\). 
This formula is rewritten by 
\begin{equation}
\Pi_{\beta} 
= 
\frac{1 + \sigma^{3}}{2} \otimes \Pi_{\beta}^{+}
+
\frac{1 - \sigma^{3}}{2} \otimes \Pi_{\beta}^{-} .
\end{equation} 
The matrix $\Pi_{\beta}^{+}$ acts on $H^{+}_{{\rm eff},\,\beta}$. 
It indicates that in the projected space the spatial inversion leads to 
\(
H^{+}_{{\rm eff},\,\beta} (\veck)
\to 
(\Pi_{\beta}^{+})^{\dagger}
H^{+}_{{\rm eff},\,\beta} (\veck)
(\Pi_{\beta}^{+})
\). 
A straightforward calculation of $\Pi_{\beta}$ leads to the fact that
$\Pi_{\beta}^{+}$ are the identity matrix. 




\begin{thebibliography}{99}
\bibitem{Hor;Cava:2010}
Y. S. Hor, A. J. Williams, J. G. Checkelsky, P. Roushan, J. Seo, Q. Xu,
H. W. Zandbergen, A. Yazdani, N. P. Ong, R. J. Cava, 
Phys. Rev. Lett. \textbf{104} (2010) 05700. 
\bibitem{Wray;Hasan:2010}
L. A. Wray, S.-Y. Xu, Y. Xia, Y. S. Hor, D. Qian, A. V. Fedorov, H. Lin,
	A. Bansil, R. J. Cava, M. Z. Hasan, 
Nat. Phys. \textbf{6} (2010) 855. 
\bibitem{Shirasawa;Takahashi:2014}
T. Shirasawa, M. Sugiki, T. Hirahara, M. Aitani, T. Shirai, S. Hasegawa,
	T. Takahashi, 
Phys. Rev. B \textbf{89} (2014) 195311. 
\bibitem{Sasaki;Ando:2014}
S. Sasaki, K. Segawa, Y. Ando, 
Phys. Rev. B \textbf{90} (2014) 220504(R). 
\bibitem{Sasaki;Ando:2011}
S. Sasaki, M. Kriener, K. Segawa, K. Yada, Y. Tanaka, M. Sato, Y. Ando,  
Phys. Rev. Lett. \textbf{107} (2011) 217001.
\bibitem{Kirzhner;Kanigel:2012}
T. Kirzhner, E. Lahoud, K. B. Chaska, Z. Salman, A. Kanigel, 
Phys. Rev. B \textbf{86} (2012) 064517. 
\bibitem{Das;Kadowaki:2011}
P. Das, Y. Suzuki, M. Tachiki, K. Kadowaki, 
Phys. Rev. B \textbf{83} (2011) 220513(R). 
\bibitem{Levy;Stroscio:2013}
N. Levy, T. Zhang, J. Ha, F. Sharifi, A. Alec Talin, Y. Kuk, J. A. Stroscio, 
Phys. Rev. Lett. \textbf{110} (2013) 117001 (2013).  
\bibitem{Hashimto;Tanaka:2012}
T. Hashimoto, K. Yada, A. Yamakage, M. Sato, Y. Tanaka, 
J. Phys. Soc. Jpn. \textbf{82} (2012) 044704.
\bibitem{Balatsky;Zhu:2006}
A. V. Balatsky, I. Vekhter, J.-X. Zhu, 
Rev. Mod. Phys. \textbf{78} (2006) 373. 
\bibitem{Nagai;Machida:2014}
Y. Nagai, Y. Ota, M. Machida, 
Phys. Rev. B \textbf{89} (2014) 214506. 
\bibitem{Hao;Lee:2011}
L. Hao, T. K. Lee, 
Phys. Rev. B \textbf{83} (2011) 134516. 
\bibitem{Fu;Berg:2010}
L. Fu, E. Berg, 
Phys. Rev. Lett. \textbf{105} (2010) 097001. 
\bibitem{Mizushima;Tanaka:2014}
T. Mizushima, A. Yamakage, M. Sato, Y. Tanaka, 
Phys. Rev. B \textbf{90} (2014) 184516. 
\bibitem{Sato:2010}
M. Sato, 
Phys. Rev. B \textbf{81} (2010) 220504(R). 
\bibitem{Nagai;1410.4646}
Y. Nagai, 
Phys. Rev. B \textbf{91} (2015) 060502(R).
\bibitem{Michaeli;Fu:2012}
K. Michaeli, L. Fu, 
Phys. Rev. Lett. \textbf{109} (2012) 187003. 
\bibitem{Alicea;Fisher:2011}
J. Alicea, Y. Oreg, G. Refael, F. von Oppen, M. P. A. Fisher, 
Nat. Phys. \textbf{7} (2011) 412. 
\bibitem{Hao;Lee:2015}
L. Hao and T.--K. Lee, 
J. Phys.: Condens. Matter \textbf{27} (2015) 105701. 
\bibitem{Hubac;Wilson:2010}
I. Huba\v{c}, S. Wilson, 
Brillouin-Wigner Methods for Many-Body Systems, 
Springer, Heidelberg, 2010.
\end{thebibliography}


\end{document}